\documentclass[aps,prd,showpacs,superscriptaddress,notitlepage,nofootinbib]{revtex4-1}
\usepackage{amsmath}
\usepackage{epsfig}
\usepackage{multirow}
\def\bea#1\eea{\begin{align}#1\end{align}}
\newcommand{\nnu}{\nonumber\\}
\newcommand{\bef}{\begin{figure}[h!tb]\centering}
\newcommand{\eef}{\end{figure}}
\allowdisplaybreaks

\begin{document}
\title{Spin asymmetries for vector boson production in polarized p+p collisions}

\author{Jin Huang}
\email{jhuang@bnl.gov}
\affiliation{Physics Department,
                   Brookhaven National Laboratory,
                   Upton, New York 11973, USA}

\author{Zhong-Bo Kang}
\email{zkang@lanl.gov}
\affiliation{Theoretical Division,
                   Los Alamos National Laboratory,
                   Los Alamos, New Mexico 87545, USA}
                   
\author{Ivan Vitev}
\email{ivitev@lanl.gov}
\affiliation{Theoretical Division,
                   Los Alamos National Laboratory,
                   Los Alamos, New Mexico 87545, USA}                   

\author{Hongxi Xing}
\email{hxing@lanl.gov}
\affiliation{Theoretical Division,
                   Los Alamos National Laboratory,
                   Los Alamos, New Mexico 87545, USA}
\date{\today}         

\begin{abstract}
We study the cross section for vector boson ($W^{\pm}/Z^0/\gamma^*$) production in polarized nucleon-nucleon collisions for low transverse momentum of the observed vector boson. For the case where one measures the transverse momentum and azimuthal angle of the vector bosons, we present the cross sections and the associated spin asymmetries in terms of transverse momentum dependent parton distribution functions (TMDs) at tree level within the TMD factorization formalism. To assess the feasibility of experimental measurements, we estimate the spin asymmetries for $W^{\pm}/Z^0$ boson production in polarized proton-proton collisions at the Relativistic Heavy Ion Collider (RHIC) by using current knowledge of the relevant TMDs. We find that some of these asymmetries can be sizable if the suppression effect from TMD evolution is not too strong. The $W$ program at RHIC can, thus, test and constrain spin theory by providing unique information on the universality properties of TMDs, TMD evolution, and the nucleon structure. For example, the single transverse spin asymmetries could be used to probe the well-known Sivers function $f_{1T}^{\perp q}$, as well as the transversal helicity distribution $g_{1T}^{q}$ via the parity-violating nature of $W$ production.
\end{abstract}

\date{\today}

\maketitle
%%%%%%%%%%
\section{Introduction}
Spin-dependent observables for  vector boson ($W^{\pm}/Z^0$) production in polarized nucleon-nucleon collisions offer excellent sensitivity to the spin-dependent parton distribution functions of the individual parton flavors in the nucleon~\cite{Aschenauer:2013woa,Aschenauer:2015eha}. For example, the violation of parity in the weak interactions gives rise to the single longitudinal spin asymmetries for $W^{\pm}$ production in proton-proton collisions~\cite{Bourrely:1993dd,Bourrely:1994sc}. Such {\it longitudinal} spin asymmetries provide flavor separation for the quark helicity distributions for $u,~\bar u,~d,~\bar d$, and, in particular, probe the antiquark polarization. For recent theoretical work and experimental measurements, see e.g. Refs.~\cite{deFlorian:2010aa,deFlorian:2011ia,Kang:2011qz,Ringer:2015oaa} and \cite{Aggarwal:2010vc,Adare:2010xa,Adamczyk:2014xyw,Adare:2015gsd,Aidala:2012mv}, respectively.

Single {\it transverse} spin asymmetries for vector bosons have also been proposed to probe the quark Sivers functions $f_{1T}^{\perp q}$ of individual flavors~\cite{Brodsky:2002pr,Schmidt:2003wi,Kang:2009bp,Kang:2009sm}. Quark Sivers functions represent the distributions of unpolarized quarks inside a transversely polarized nucleon through a correlation between the transverse momentum of the quark and the transverse spin of the nucleon, and they have a unique time-reversal modified universality property. It was shown from the parity and time-reversal invariance of QCD that the quark Sivers functions in semi-inclusive deep inelastic scattering (SIDIS) and those in Drell-Yan (DY) type process (e.g., $W^{\pm}/Z^0,~\gamma^*$) should have the same functional form but an opposite sign -- so-called ``sign change''~\cite{Collins:2002kn,Boer:2003cm,Brodsky:2002cx,Brodsky:2002rv,Kang:2009bp}. Quark Sivers functions have been measured/extracted from SIDIS process, see, e.g. \cite{Anselmino:2008sga,Gamberg:2013kla}. So far mainly the valence $u$ and $d$ quark Sivers functions were constrained, while sea quark Sivers functions remain largely unknown~\cite{Echevarria:2014xaa}. Single transverse spin asymmetries for vector bosons in proton-proton collisions can thus serve as these purposes, i.e., to test the sign change on one side, and to constrain the sea quark Sivers function at the same time. The sea quark distribution will be further measured with high precision in a future electron ion collider~\cite{Accardi:2012qut,Aschenauer:2014cki,Abeyratne:2015pma}.

There are also theoretical studies of double spin asymmetries for vector boson production. For example, double transverse spin asymmetries for Drell-Yan production were originally proposed to extract the quark transversity distributions~\cite{Ralston:1979ys,Artru:1989zv,Jaffe:1991ra,Cortes:1991ja,Boer:2000er}. However, as pointed out already in~\cite{Boer:2011vq}, there could be additional terms involving the product of quark Sivers function $f_{1T}^{\perp q} f_{1T}^{\perp \bar q}$ (called double Sivers effect), as well as the product $g_{1T}^{q} g_{1T}^{\bar q}$ (called double worm-gear effect in~\cite{Boer:2011vq} since $g_{1T}^{q}$ is sometimes referred to as a worm-gear function). There are, of course, longitudinal-transverse double spin asymmetries for the vector boson production, which could provide complementary information on the quark helicity distribution functions~\cite{Metz:2010xs,Metz:2012fq}.

The function $g_{1T}^{q}$ is also called ``transversal helicity'' distribution function~\cite{Mulders:1995dh,Bacchetta:2006tn,Boer:1997nt}, as it gives the quark longitudinal polarization inside a transversely polarized proton. $g_{1T}^{q}$ could also appear in single transverse spin asymmetry of $W^{\pm}$ production due to the parity-violating nature of the weak interaction, see e.g.  a study within the collinear twist-3 formalism~\cite{Metz:2010xs}. Contrary to the quark Sivers function $f_{1T}^{\perp q}$ that changes sign from SIDIS to DY type processes, the transversal helicity distribution $g_{1T}^{q}$ is universal between SIDIS and DY. As $g_{1T}^{q}$ has been investigated in the SIDIS measurements~\cite{Parsamyan:2007ju,Pappalardo:2011cu,Huang:2011bc,Parsamyan:2015dfa} and is proposed to be measured with high precision in future SIDIS experiments~\cite{Gao:2010av}, one in principle could test the universality of $g_{1T}^{q}$ in $W^{\pm}/Z^0$ production in proton-proton collisions at RHIC. 

RHIC has planned a dedicated transversely polarized proton-proton  run to measure the spin asymmetries for  $W^{\pm}$ production~\cite{Aschenauer:2015eha}. Together with its longitudinal $W^{\pm}$ physics, this will ensure a comprehensive $W^{\pm}$ physics spin program. Besides the sensitivity to the spin-dependent parton distribution functions inside the nucleon, vector boson production also provides excellent  constraints on the QCD evolution effects. This is because the mass of the vector boson, which sets the typical scale of the hard scattering, is usually much larger than the scales probed in the fixed-target SIDIS measurements. It is, thus, timely to present in a single, self-contained paper the results for the spin asymmetries for vector boson production in polarized nucleon-nucleon collisions. We will consider the production of  vector bosons at small transverse momentum, and thus a transverse momentum dependent (TMD) factorization formalism is the appropriate framework~\cite{Collins:1981uk,Ji:2004wu,Ji:2004xq}. Since it is now possible to perform a full reconstruction of the produced boson kinematics~\cite{Adamczyk:2015gyk,Aschenauer:2015eha}, we will integrate out the kinematics of the decayed leptons, and present the results at the level of the vector boson.  This will simplify the physics results to facilitate the interpretation of the experimental measurements. 

The rest of the paper is organized as follows. In Sec.~\ref{sec:II}, we present the TMD formalism for all  spin asymmetries for vector boson production ($W^{\pm}/Z^0, ~\gamma^*$) in polarized nucleon-nucleon collisions, where one measures the transverse momentum $q_T$ and azimuthal angle $\phi_V$ of the vector bosons in the center-of-mass frame of the colliding nucleons. We point out that due to the parity-violation nature of the weak interaction for $W^{\pm}/Z^0$ production, besides the well-known Sivers term $\propto \sin(\phi_V-\phi_S)$, the single transverse spin asymmetry can have an additional term $\propto \cos(\phi_V - \phi_S)$ with $\phi_S$ the azimuthal angle of the transverse spin of the nucleon, which is related to the transversal helicity distribution $g_{1T}^q$~\cite{Metz:2010xs}. In Sec.~\ref{sec:III} we present the numerical estimate of these single and double spin asymmetries for polarized proton-proton collisions at RHIC and find that some asymmetries are sizable if TMD evolution does not lead to too strong a suppression.  We conclude our paper in Sec.~\ref{sec:IV}. 

%%%%%%%%%%
\section{Spin-dependent cross section for vector boson production}
\label{sec:II}
In this section, we derive the spin-dependent differential cross section for vector boson production ($V=W^{\pm}$, $Z^0$, or $\gamma^*$) in polarized nucleon-nucleon scattering 
\bea
A(P_A, S_A)+B(P_B, S_B) \rightarrow V(y, q_T, \phi_V)+X,
\eea
where a polarized nucleon $A$ of momentum $P_A$ and spin $S_A$ is colliding with another nucleon $B$ of momentum $P_B$ and spin $S_B$. We work in the center-of-mass (CM) frame of the colliding nucleons with $S=(P_A+P_B)^2$ the CM energy squared. In such a frame, we choose the nucleon $A$ to be moving in the ``$+z$'' direction, while the nucleon $B$ is moving in the ``$-z$'' direction. In the final state, we have the full kinematics of the vector boson -- its rapidity $y$ and transverse momentum, with magnitude $q_T$ and azimuthal angle $\phi_V$.  For a virtual photon we further observe its invariant mass $Q$. We will concentrate on the kinematic region where the transverse momentum $q_T$ is much less than the mass $M_V$ of the vector boson: $q_T \ll M_V$. This is the region where the usual TMD factorization formalism is expected to be valid~\cite{Collins:1981uk,Ji:2004wu,Ji:2004xq}. 

We will take $W$ boson production as an example to work out the leading order differential cross section in terms of the TMDs. The derivation for $Z$ boson and virtual photon will be similar, and the corresponding results will be presented at the end of this section. The differential cross section for $W$ boson production can be written as
\bea
\frac{d\sigma^W}{d^4q} = \frac{1}{2S}\,\left(\frac{g_W}{2\sqrt{2}}\right)^2\, \sum_\lambda \epsilon^\lambda_{\mu}(q) \epsilon^{*\lambda}_{\nu}(q)\, W^{\mu\nu}\left(P_A, S_A, P_B, S_B\right)\,
2\pi\delta\left(q^2 - M_W^2\right),
\label{eq-xsec1}
\eea
where $g_W$ is the weak charge and related to the electric charge as $e=g_W \sin\theta_W$ with $\theta_W$ the Weinberg angle. Using
\bea
&\frac{g_W^2}{8 M_W^2} = \frac{G_F}{\sqrt{2}},\qquad
d^4q=\frac{1}{2} dq^2\, dy d^2q_\perp,  
\\
&\sum_\lambda \epsilon^\lambda_{\mu}(q) \epsilon^{*\lambda}_{\nu}(q) = -g_{\mu\nu} + \frac{q_\mu q_\nu}{M_W^2},
\label{spin-sum}
\eea
with $G_F$ the Fermi constant, we obtain
\bea
\frac{d\sigma^W}{dy d^2\vec q_T} = \frac{\pi G_F M_W^2}{2\sqrt{2} S} \left(-g_{\mu\nu} + \frac{q_\mu q_\nu}{M_W^2}\right)
W^{\mu\nu}\left(P_A, S_A, P_B, S_B\right),
\eea
where $W^{\mu\nu}$ is the hadronic tensor and is given by
\bea
W^{\mu\nu}\left(P_A, S_A, P_B, S_B\right) =& \frac{1}{N_c} \sum_{q,q'} \left|V_{qq'}\right|^2 \int d^2\vec k_{aT}d^2\vec k_{bT}\delta^2\left(\vec q_T-\vec k_{aT}-\vec k_{bT}\right) 
\nnu
&\times {\rm Tr}\left[\gamma^{\mu}(v_q-a_q\gamma^5)\Phi^q(x_a,\vec k_{aT}, S_{A})\gamma^{\nu}(v_q-a_q\gamma^5)\bar\Phi^{q'}(x_b,\vec k_{bT}, S_{B})\right].
\eea
Here $N_c=3$ is the number of colors, $V_{qq'}$ are the CKM elements for weak interaction, $v_q$ ($a_q$) is the vector (axial) coupling of the $W$ boson to the quark with 
\bea
v_q = 1, \qquad
a_q = 1. 
\eea
The transverse momentum dependent quark-quark correlators are defined as~\cite{Arnold:2008kf}
\bea
\Phi^q(x_a, \vec k_{aT}, S_{A}) &= \int \frac{dz^- d^2 \vec{z}_T}{\left(2\pi\right)^3}e^{ik_a^+z^- - i\vec{k}_{aT}\cdot \vec{z}_T} \langle P_A, S_A| \bar\psi_j^q(0) \psi_i^q(z)| P_A, S_A\rangle,
\\
\bar{\Phi}^q(x_b, \vec k_{bT}, S_{B}) &= \int \frac{dz^+ d^2 \vec{z}_T}{\left(2\pi\right)^3}e^{ik_b^-z^+ - i\vec{k}_{bT}\cdot \vec{z}_T} \langle P_B, S_B| \psi_i^q(0) \bar \psi_j^q(z)| P_B, S_B\rangle,
\eea
where we have suppressed the gauge link dependence in the definitions. They can be expanded as the following parametrization in terms of the Dirac matrices at leading-twist
\bea
\Phi^q(x_a,\vec k_{aT}, S_{A}) = \Phi^{q[\gamma^+]}\frac{\gamma^-}{2} + \Phi^{q[\gamma^+\gamma^5]}\frac{\gamma^5\gamma^-}{2}
+\Phi^{q[i\sigma^{\alpha+}\gamma^5]}\frac{-i\sigma^{\alpha-}\gamma^5}{2},
\label{eq-phiq}
\eea
where $\alpha=\{1,2\}$ is a transverse index, $\Phi^{q[\Gamma]}\equiv \frac{1}{2}{\rm Tr}[\Phi^q\Gamma]$ for the specific Dirac matrix $\Gamma$ as given above, and likewise for $\bar{\Phi}^q$. With such an expansion at hand, we thus have the $W$ boson cross section as
\bea
\frac{d\sigma^W}{dy d^2\vec q_T} =& \sigma_0^W \sum_{q,q'} \left|V_{qq'}\right|^2 \int d^2\vec k_{aT}d^2\vec k_{bT}\delta^2\left(\vec q_T-\vec k_{aT}-\vec k_{bT}\right) \Bigg[(v_q^2+a_q^2)  \left(\Phi^{q[\gamma^+]} \bar\Phi^{q'[\gamma^-]} \right.
\nnu
&\left.
+\Phi^{q[\gamma^+\gamma^5]} \bar\Phi^{q'[\gamma^-\gamma^5]}\right)
-2a_q v_q\left(\Phi^{q[\gamma^+]} \bar\Phi^{q'[\gamma^-\gamma^5]}
+\Phi^{q[\gamma^+\gamma^5]}\bar\Phi^{q'[\gamma^-]}\right) +\left(\Phi\leftrightarrow \bar \Phi\right)\Bigg],
\label{eq-xsec2}
\eea
where $\sigma_0^W$ is given by
\bea
\sigma_0^W = \frac{\pi G_F M_W^2}{\sqrt{2} S N_c}.
\eea
It is instructive to note that there are no terms involving $\Phi^{q[\Gamma]}$ with $\Gamma = i\sigma^{\alpha\pm}\gamma^5$, i.e. the last term in Eq.~\eqref{eq-phiq} does not contribute to the final result. These terms are related to the transversely polarized quark distributions inside the proton. They contribute to the cross section only when one measures the angular dependence of the leptons decayed from the vector boson. When we integrate over the phase space of the decayed leptons, i.e. when one sums over the spin states of the vector boson, and contracts $W^{\mu\nu}$ with $\sum_\lambda \epsilon^\lambda_{\mu}(q) \epsilon^{*\lambda}_{\nu}(q)$ as specified in Eq.~\eqref{spin-sum}, they vanish. 

We can further express the cross section in terms of leading-twist quark TMDs through the following standard expressions of $\Phi^{q[\Gamma]}$ for quark~\cite{Arnold:2008kf}:
\bea
\Phi^{q[\gamma^+]} &= f_1^q(x_a,\vec k_{aT}^2) - \frac{\epsilon_T^{ij}k_{aT}^i S_{AT}^j}{M_A}f_{1T}^{\perp q}(x_a,\vec k_{aT}^2),
\label{eq-gamma+}
\\
\Phi^{q[\gamma^+\gamma^5]} &= S_{AL}g_{1L}^q(x_a,\vec k_{aT}^2)+\frac{\vec k_{aT}\cdot \vec S_{AT}}{M_A}g_{1T}^q(x_a,\vec k_{aT}^2),
\label{eq-gamma+5}
\eea
where $f_1^q$ is the unpolarized quark TMD, and $f_{1T}^{\perp q}$ is the so-called quark Sivers function~\cite{Sivers:1989cc,Sivers:1990fh}. $g_{1L}^q$ is the quark helicity distribution function, describing the quark longitudinal polarization in a longitudinally polarized proton. Finally, as we have mentioned already in the Introduction, $g_{1T}^q$ is the ``transversal helicity'' distribution function, describing the quark longitudinal polarization in a transversely polarized proton~\cite{Mulders:1995dh,Bacchetta:2006tn,Boer:1997nt}. Similarly we have the expressions for $\bar\Phi^{q[\gamma^-]}$ for the antiquark: 
\bea
\bar\Phi^{q[\gamma^-]} &= f_1^{\bar q}(x_b,\vec k_{bT}^2) + \frac{\epsilon_T^{ij}k_{bT}^i S_{BT}^j}{M_B}f_{1T}^{\perp \bar q}(x_b,\vec k_{bT}^2),\\
\bar \Phi^{q[\gamma^- \gamma^5]} &= -S_{BL}g_{1L}^{\bar q}(x_b,\vec k_{bT}^2)-\frac{\vec k_{bT}\cdot \vec S_{BT}}{M_B}g_{1T}^{\bar q}(x_b,\vec k_{bT}^2).
\eea
Here $M_A$ ($M_B$) are the mass of the nucleon $A$ ($B$), while $S_{AL},~\vec S_{AT}$ and $S_{BL},~\vec S_{BT}$ are the longitudinal and transverse components of the spin vector for nucleon $A$ and $B$, respectively. It might be worthwhile to remind the reader that $\Phi^{q[\Gamma]}$ is associated with the incoming nucleon $A$, which moves in ``$+z$'' direction in the CM frame. On the other hand, the correlator $\bar \Phi^{q[\Gamma]}$ is associated with the incoming nucleon $B$, which moves in ``$-z$'' direction. Their spin vectors in the CM frame have the following forms:
\bea
S_A^\mu =& \left(S_{AL} \frac{|\vec P_A|}{M_A}, \, |\vec S_{AT}| \cos\phi_{S_A}, \, |\vec S_{AT}| \sin \phi_{S_A}, \, S_{AL} \frac{P_A^0}{M_A} \right),
\\
S_B^\mu =& \left(S_{BL} \frac{|\vec P_B|}{M_B}, \, |\vec S_{BT}| \cos\phi_{S_B}, \, |\vec S_{BT}| \sin \phi_{S_B}, \, -S_{BL} \frac{P_B^0}{M_B} \right). 
\eea 
Substituting the parametrization of $\Phi^{q[\Gamma]}$ and $\bar \Phi^{q[\Gamma]}$ into Eq.~\eqref{eq-xsec2}, we arrive at the final result:
\bea
\frac{d\sigma^W}{dyd^2\vec q_T} =& \sigma_0^W  \Bigg\{
F_{UU} + S_{AL} F_{LU} + S_{BL} F_{UL} + S_{AL} S_{BL} F_{LL}
\nnu
&+ |\vec S_{AT}| \left[\sin(\phi_V - \phi_{S_A}) F_{TU}^{\sin(\phi_V-\phi_{S_A})}
+ \cos(\phi_V - \phi_{S_A}) F_{TU}^{\cos(\phi_V - \phi_{S_A})}
\right]
\nnu
&+ |\vec S_{BT}| \left[\sin(\phi_V - \phi_{S_B}) F_{UT}^{\sin(\phi_V - \phi_{S_B})}
+ \cos(\phi_V - \phi_{S_B}) F_{UT}^{(\cos\phi_V - \phi_{S_B})}
\right]
\nnu
&+ |\vec S_{AT}| S_{BL} \left[ \sin(\phi_V - \phi_{S_A}) F_{TL}^{\sin(\phi_V - \phi_{S_A})}
+ \cos(\phi_V - \phi_{S_A}) F_{TL}^{\cos(\phi_V - \phi_{S_A})}
\right]
\nnu
&+ S_{AL} |\vec S_{BT}| \left[ \sin(\phi_V - \phi_{S_B}) F_{LT}^{\sin(\phi_V - \phi_{S_B})}
+ \cos(\phi_V - \phi_{S_B}) F_{LT}^{\cos(\phi_V - \phi_{S_B})}
\right]
\nnu
&+ |\vec S_{AT}| |\vec S_{BT}| \left[
\cos(2\phi_V - \phi_{S_A} - \phi_{S_B}) F_{TT}^{\cos(2\phi_V - \phi_{S_A} - \phi_{S_B})}
+\cos(\phi_{S_A}-\phi_{S_B}) F_{TT}^1\right.
\nnu
&\hspace{19mm}+\left.\sin(2\phi_V - \phi_{S_A} - \phi_{S_B}) F_{TT}^{\sin(2\phi_V - \phi_{S_A} - \phi_{S_B})}
+\sin(\phi_{S_A}-\phi_{S_B}) F_{TT}^2\right]\Bigg\}.
\label{eq:W}
\eea
To write explicitly the expressions for all the structure functions $F$ in the above equation, let us define the following short-hand notation:
\bea
{\mathcal C}^W\left[w(\vec k_{aT}, \vec k_{bT}) f_1\bar f_2\right] =&  \sum_{q,q'} \left|V_{qq'}\right|^2 \int d^2\vec k_{aT}d^2\vec k_{bT} \delta^2\left(\vec q_T-\vec k_{aT}-\vec k_{bT}\right) w(\vec k_{aT}, \vec k_{bT})
\nnu
&\times \left[f_1^q(x_a, \vec k_{aT}^2) f_2^{\bar q'}(x_b, \vec k_{bT}^2) + \left(q\leftrightarrow \bar q'\right) \right].
\label{eq:CW}
\eea
One should be careful in the second term $q\leftrightarrow \bar q'$ when we interchange the roles of quarks and antiquarks. Due to the fact that~\cite{Tangerman:1994eh,Arnold:2008kf}
\bea
\bar \Phi^{q[\Gamma]} = \pm \Phi^{\bar q[\Gamma]}, \qquad
\begin{cases}
+ & {\rm for~} \gamma^\mu,~i\sigma^{\mu\nu} \gamma^5 \\
- & {\rm for~} \gamma^\mu\gamma^5,~1,~i\gamma^5
\end{cases}
\, ,
\eea
when a term involves an odd number of $g_{1L}$ and/or $g_{1T}$, there should be a minus sign when one interchanges $q\leftrightarrow \bar q'$. For example,
\bea
{\mathcal C}^W \left[ (v_q^2+a_q^2)\, f_1 \, \bar f_1\right] = &  \sum_{q,q'} \left|V_{qq'}\right|^2 \int d^2\vec k_{aT}d^2\vec k_{bT} \delta^2\left(\vec q_T-\vec k_{aT}-\vec k_{bT}\right) (v_q^2+a_q^2)
\nnu
&\times \left[f_1^q(x_a, \vec k_{aT}^2) f_1^{\bar q'}(x_b, \vec k_{bT}^2) + f_1^{\bar q'}(x_a, \vec k_{aT}^2) f_1^{q}(x_b, \vec k_{bT}^2)\right],
\label{eq:f1f1}
\\
{\mathcal C}^W \left[ 2 v_q a_q
\frac{\hat q_T\cdot\vec k_{aT}}{M_A}  g_{1T} \, \bar f_1\right]
= &  \sum_{q,q'} \left|V_{qq'}\right|^2 \int d^2\vec k_{aT}d^2\vec k_{bT} \delta^2\left(\vec q_T-\vec k_{aT}-\vec k_{bT}\right) 2 v_q a_q \frac{\hat q_T\cdot\vec k_{aT}}{M_A}
\nnu
&\times \left[g_{1T}^q(x_a, \vec k_{aT}^2) f_1^{\bar q'}(x_b, \vec k_{bT}^2) - g_{1T}^{\bar q'}(x_a, \vec k_{aT}^2) f_1^{q}(x_b, \vec k_{bT}^2)\right],
\label{eq:g1tf1}
\\
{\mathcal C}^W \left[ (v_q^2+a_q^2) 
\frac{\hat q_T\cdot\vec k_{aT}}{M_A}   g_{1T} \, \bar g_{1L} \right]
= & \sum_{q,q'} \left|V_{qq'}\right|^2 \int d^2\vec k_{aT}d^2\vec k_{bT} \delta^2\left(\vec q_T-\vec k_{aT}-\vec k_{bT}\right) (v_q^2+a_q^2) \frac{\hat q_T\cdot\vec k_{aT}}{M_A} 
\nnu
&\times \left[g_{1T}^q(x_a, \vec k_{aT}^2) g_{1L}^{\bar q'}(x_b, \vec k_{bT}^2) + g_{1T}^{\bar q'}(x_a, \vec k_{aT}^2) g_{1L}^{q}(x_b, \vec k_{bT}^2)\right].
\label{eq:g1tg1l}
\eea
Here the first and the third equations, Eqs.~\eqref{eq:f1f1} and \eqref{eq:g1tg1l}, do not have a minus sign under $q\leftrightarrow \bar q'$. This is because they either  do not involve any $g_{1T}$ or $g_{1L}$ at all as in Eq.~\eqref{eq:f1f1}, or they involve even total number of $g_{1T}$ and $g_{1L}$ as in Eq.~\eqref{eq:g1tg1l}. On the other hand, since Eq.~\eqref{eq:g1tf1} involves an odd total number of $g_{1T}$ or $g_{1L}$, there is a minus sign under $q\leftrightarrow \bar q'$.

Using this shorthand notation and defining $\hat q_T = \vec q_T/ q_T$, we have
\bea
F_{UU} =& {\mathcal C}^W \left[ (v_q^2+a_q^2)\, f_1 \, \bar f_1\right],
\\
F_{LU} =& - {\mathcal C}^W \left[ 2 v_q a_q \, g_{1L} \, \bar f_1 \right],
\\
F_{UL} =&  {\mathcal C}^W \left[ 2 v_q a_q \, f_1 \, \bar g_{1L} \right],
\\
F_{LL} =& - {\mathcal C}^W \left[ (v_q^2+a_q^2)\, g_{1L} \, \bar g_{1L} \right],
\\
F_{TU}^{\sin(\phi_V - \phi_{S_A})} =& {\mathcal C}^W \left[ (v_q^2+a_q^2) 
\frac{\hat q_T\cdot\vec k_{aT}}{M_A}  f_{1T}^{\perp} \, \bar f_1\right],
\\
F_{TU}^{\cos(\phi_V - \phi_{S_A})} =& - {\mathcal C}^W \left[ 2 v_q a_q
\frac{\hat q_T\cdot\vec k_{aT}}{M_A}  g_{1T} \, \bar f_1\right],
\label{eq:ftu_cos}
\\
F_{UT}^{\sin(\phi_V - \phi_{S_B})} =& - {\mathcal C}^W \left[ (v_q^2+a_q^2) 
\frac{\hat q_T\cdot\vec k_{bT}}{M_B} f_1 \, \bar f_{1T}^{\perp} \right],
\\
F_{UT}^{\cos(\phi_V - \phi_{S_B})} =& {\mathcal C}^W \left[ 2 v_q a_q
\frac{\hat q_T\cdot\vec k_{bT}}{M_B}   f_1 \, \bar g_{1T} \right],
\label{eq:fut_cos}
\\
F_{TL}^{\sin(\phi_V - \phi_{S_A})} =& {\mathcal C}^W \left[ 2 v_q a_q
\frac{\hat q_T\cdot\vec k_{aT}}{M_A}   f_{1T}^{\perp} \, \bar g_{1L} \right],
\\
F_{TL}^{\cos(\phi_V - \phi_{S_A})} =& - {\mathcal C}^W \left[ (v_q^2+a_q^2) 
\frac{\hat q_T\cdot\vec k_{aT}}{M_A}   g_{1T} \, \bar g_{1L} \right],
\\
F_{LT}^{\sin(\phi_V - \phi_{S_B})} =&  {\mathcal C}^W \left[ 2 v_q a_q
\frac{\hat q_T\cdot\vec k_{bT}}{M_B}  g_{1L}  \, \bar f_{1T}^{\perp}  \right],
\\
F_{LT}^{\cos(\phi_V - \phi_{S_B})} =& - {\mathcal C}^W \left[ (v_q^2+a_q^2) 
\frac{\hat q_T\cdot\vec k_{bT}}{M_B}   g_{1L} \, \bar g_{1T} \right],
\\
F_{TT}^{\cos(2\phi_V - \phi_{S_A} - \phi_{S_B})} =& {\mathcal C}^W \left[ 
(v_q^2+a_q^2)
\frac{2\vec k_{aT}\cdot \hat q_T \vec k_{bT}\cdot \hat q_T-\vec k_{aT}\cdot \vec k_{bT}}{2M_AM_B}
(f_{1T}^{\perp} \bar f_{1T}^{\perp} - g_{1T} \bar g_{1T})  \right],
\\
F_{TT}^{\sin(2\phi_V - \phi_{S_A} - \phi_{S_B})} =& {\mathcal C}^W \left[ 
v_q a_q
\frac{2 \vec k_{aT}\cdot \hat q_T \vec k_{bT}\cdot \hat q_T-\vec k_{aT}\cdot \vec k_{bT}}{M_AM_B}
(f_{1T}^{\perp} \bar g_{1T} + g_{1T} \bar f_{1T}^{\perp})  \right],
\\
F_{TT}^{1} =& - {\mathcal C}^W \left[ 
(v_q^2+a_q^2)\frac{\vec k_{aT}\cdot \vec k_{bT}}{2M_AM_B} (f_{1T}^{\perp} \bar f_{1T}^{\perp}+g_{1T} \bar g_{1T}) \right],
\\
F_{TT}^{2} =& - {\mathcal C}^W \left[ 
v_q a_q \frac{\vec k_{aT}\cdot \vec k_{bT}}{M_AM_B} (f_{1T}^{\perp}\bar g_{1T} - g_{1T} \bar f_{1T}^{\perp}) \right].
\eea

To obtain the differential cross section for $Z$ boson production in polarized proton-proton collisions, one simply replaces $\sigma_0^W$ by $\sigma_0^Z$ in Eq.~\eqref{eq:W} with 
\bea
\sigma_0^Z = \frac{\sqrt{2} \pi G_F M_Z^2}{S N_c}.
\eea
All the structure functions will be given by the same expressions as above, with ${\mathcal C}^W$ in Eq.~\eqref{eq:CW} replaced by ${\mathcal C}^Z$:
\bea
{\mathcal C}^Z\left[w(\vec k_{aT}, \vec k_{bT})f_1\bar f_2\right] =  \sum_{q} \int d^2\vec k_{aT}d^2\vec k_{bT}\delta^2\left(\vec q_T-\vec k_{aT}-\vec k_{bT}\right) w(\vec k_{aT}, \vec k_{bT})
\left[f_1^q(x_a, \vec k_{aT}^2) f_2^{\bar q}(x_b, \vec k_{bT}^2) + \left(q\leftrightarrow \bar q\right)  \right].
\label{eq:CZ}
\eea
At the same time, for $Z$ bosons we have
\bea
v_q=T^3_q - 2 e_q \sin^2\theta_W, \qquad a_q=T^3_q,
\eea
where $e_q$ is the quark electric charge and $T^3_q$ is the third component of ``weak isospin'' of the quark.  $T^3_q=\frac{1}{2}$  for $u,~c,~t$ quarks, while $T^3_q=-\frac{1}{2}$ for $d,~s,~b$ quarks.

Furthermore, we could also write down the differential cross section for the Drell-Yan lepton pair production through the virtual photon decay, which gives
\bea
\frac{d\sigma^{\rm DY}}{dQ^2 dyd^2\vec q_T} =& \sigma_0^{\rm DY}  \Bigg\{
F_{UU} + S_{AL} S_{BL} F_{LL}
\nnu
&+ |\vec S_{AT}| \left[\sin(\phi_V - \phi_{S_A}) F_{TU}^{\sin(\phi_V - \phi_{S_A})} \right]
+ |\vec S_{BT}| \left[\sin(\phi_V - \phi_{S_B}) F_{UT}^{\sin(\phi_V - \phi_{S_B})}\right]
\nnu
&+ |\vec S_{AT}| S_{BL} \left[\cos(\phi_V - \phi_{S_A}) F_{TL}^{\cos(\phi_V - \phi_{S_A})}\right]
+ S_{AL} |\vec S_{BT}| \left[ \cos(\phi_V - \phi_{S_B}) F_{LT}^{\cos(\phi_V - \phi_{S_B})}\right]
\nnu
&+ |\vec S_{AT}| |\vec S_{BT}| \left[
\cos(2\phi_V - \phi_{S_A} - \phi_{S_B}) F_{TT}^{\cos(2\phi_V - \phi_{S_A} - \phi_{S_B})}
+\cos(\phi_{S_A}-\phi_{S_B}) F_{TT}^1 \right]\Bigg\}.
\label{eq:DY}
\eea
Here $\sigma_0^{\rm DY}$ is given by~\cite{Echevarria:2014xaa}
\bea
\sigma_0^{\rm DY} = \frac{4\pi\alpha_{\rm em}^2}{3SQ^2N_c},
\eea
where $Q$ is the invariant mass of the lepton pair, and $\alpha_{\rm em}$ is the electromagnetic coupling constant. At the same time, all the structure functions can be obtained from those for $Z$ boson production [including the convolution expression in Eq.~\eqref{eq:CZ}] but with the replacement
\bea
v_q = e_q, \qquad a_q = 0.
\eea

Several comments are in order. 
\begin{itemize}
\item
In our setup, for  Drell-Yan production through the virtual photon channel (parity conserving channel) only the Sivers effects survive in the single transverse spin asymmetry: as can be clearly seen from Eq.~\eqref{eq:DY}, they are related to $\sin(\phi_V-\phi_{S}) $ modulation. This is because we integrate out the full kinematics of the decayed lepton pair, and only measure the azimuthal angle of the virtual photon. If instead one further measures the kinematics of the decayed lepton pair, e.g., measure both polar and azimuthal angles in the so-called Collins-Soper frame~\cite{Collins:1977iv}, one could have additional terms, such as the product of quark transversity and Boer-Mulder functions~\cite{Boer:1999mm}. For complete results in this case, see Ref. \cite{Arnold:2008kf}, as well as Ref. \cite{Pitonyak:2013dsu} where Z contribution is also included. 

\item
For $W^{\pm}/Z^{0}$ production, with parity-violating interactions there are two azimuthal modulation terms which can contribute to the single transverse spin asymmetry, as can be seen from Eq.~\eqref{eq:W}. Besides the term related to $\sin(\phi_V-\phi_{S})$ modulation just like in the Drell-Yan dilepton production, there is another term related to $\cos(\phi_V-\phi_{S})$ modulation. The $\sin(\phi_V-\phi_{S})$ term is associated with the quark Sivers function $f_{1T}^{\perp q}$, and it is the usual Sivers effect. On the other hand, the $\cos(\phi_V-\phi_{S})$ term is associated with the transversal helicity $g_{1T}^{q}$, which projects out the longitudinal quark distribution inside a transversely polarized proton. 
As in Eqs.~\eqref{eq:ftu_cos} and~\eqref{eq:fut_cos}, the relevant structure functions, $F_{TU}^{\cos(\phi_V - \phi_{S_A})}$ and $F_{UT}^{\cos(\phi_V - \phi_{S_B})}$, are directly proportional to $a_q$, and thus only exist when there is parity-violating interactions, i.e. they are unique to $W^{\pm}/Z^0$ production. The amplitudes of both $\sin(\phi_V-\phi_{S})$- and $\cos(\phi_V-\phi_{S})$ modulations can be comparable, as they involve only one spin-dependent parton distribution. Therefore, in the experimental study of the single transverse spin asymmetries for $W^{\pm}/Z^{0}$ production, one should consider extracting both terms simultaneously to avoid cross contamination, e.g. by following ``maximum-likelihood fit'' method widely used in the extraction of azimuthal spin asymmetries in the SIDIS process~\cite{Airapetian:2009ae,Alekseev:2010rw,Huang:2011bc}. 

\item
There are terms contributing to the double transverse spin asymmetries. In particular, $F_{TT}^{\cos(2\phi_V - \phi_{S_A} - \phi_{S_B})}$ receives contributions from the product of the Sivers functions $f_{1T}^{\perp q} f_{1T}^{\perp \bar q}$, as well as the product of transversal helicity distribution $g_{1T}^q g_{1T}^{\bar q}$. They were referred to in Ref.~\cite{Boer:2011vq} as double Sivers and double worm-gear effects, respectively. As pointed out already in~\cite{Boer:2011vq}, all these double transverse spin asymmetries do not involve quark transversity, contrary to the collinear factorization picture with $q_T$ integrated, in which such asymmetries usually involve the contribution from the collinear quark transversity~\cite{Ralston:1979ys,Artru:1989zv,Jaffe:1991ra,Cortes:1991ja}. 

\item
For identical nucleon scattering, e.g., proton-proton collisions at RHIC, under the exchange of the rapidity of the vector boson $y\leftrightarrow -y$, we have the following relations
\bea
&
F_{UU}(y) = F_{UU} (-y), 
\qquad
F_{LL}(y) = F_{LL} (-y), 
\qquad
F_{LU}(y) = F_{UL} (-y), 
\label{eq:asyLU}
\\
&F_{TU}^{\sin(\phi_V - \phi_{S_A})}(y) = - F_{UT}^{\sin(\phi_V - \phi_{S_B})} (-y),
\qquad
F_{TU}^{\cos(\phi_V - \phi_{S_A})}(y) =  F_{UT}^{\cos(\phi_V - \phi_{S_B})} (-y),
\label{eq:asyTU}
\\
&F_{TL}^{\sin(\phi_V - \phi_{S_A})}(y) = - F_{LT}^{\sin(\phi_V - \phi_{S_B})} (-y),
\qquad
F_{TL}^{\cos(\phi_V - \phi_{S_A})}(y) = F_{LT}^{\cos(\phi_V - \phi_{S_B})} (-y),
\label{eq:asyTL}
\\
&F_{TT}^{\cos(2\phi_V - \phi_{S_A} - \phi_{S_B})}(y) = F_{TT}^{\cos(2\phi_V - \phi_{S_A} - \phi_{S_B})}(-y),
\qquad
F_{TT}^{1}(y) = F_{TT}^{1}(-y),
\label{eq:asyTTcos}
\\
&F_{TT}^{\sin(2\phi_V - \phi_{S_A} - \phi_{S_B})}(y) = - F_{TT}^{\sin(2\phi_V - \phi_{S_A} - \phi_{S_B})}(-y),
\qquad
F_{TT}^{2}(y) = - F_{TT}^{2}(-y).
\label{eq:asyTTsin}
\eea

\end{itemize}

%%%%%%%%%%%
\section{Phenomenology at the RHIC energy}
\label{sec:III}
In this section, we move on to discuss numerical estimates for the magnitude of the spin asymmetries in polarized proton-proton collisions at the top RHIC energy. 

%%%%%%%%%%%%%%%%%
\subsection{Definitions and parametrizations}
To start, we first define various spin asymmetries for the vector boson production in polarized proton-proton collisions at RHIC. Longitudinal spin asymmetries do not involve any azimuthal angle dependence, and we define the single longitudinal spin asymmetry $A_{LU}$, $A_{UL}$ and double longitudinal spin asymmetry $A_{LL}$ as
\bea
A_{LU} = \frac{F_{LU}}{F_{UU}}, 
\qquad
A_{UL} = \frac{F_{UL}}{F_{UU}}, 
\qquad
A_{LL} = \frac{F_{LL}}{F_{UU}}.
\eea
All other spin asymmetries involve transverse spin of the incoming protons, and they can be defined similarly as follows
\bea
&A_{TU}^{\sin(\phi_V - \phi_{S_A})} = \frac{F_{TU}^{\sin(\phi_V - \phi_{S_A})}}{F_{UU}},
\qquad
A_{TU}^{\cos(\phi_V - \phi_{S_A})} = \frac{F_{TU}^{\cos(\phi_V - \phi_{S_A})}}{F_{UU}},
\\
&A_{TL}^{\sin(\phi_V - \phi_{S_A})} = \frac{F_{TL}^{\sin(\phi_V - \phi_{S_A})}}{F_{UU}},
\qquad
A_{TL}^{\cos(\phi_V - \phi_{S_A})} = \frac{F_{TL}^{\cos(\phi_V - \phi_{S_A})}}{F_{UU}},
\\
&A_{TT}^{\sin(2\phi_V - \phi_{S_A} - \phi_{S_B})} = \frac{F_{TT}^{\sin(2\phi_V - \phi_{S_A} - \phi_{S_B})}}{F_{UU}},
\qquad
A_{TT}^{\cos(2\phi_V - \phi_{S_A} - \phi_{S_B})} = \frac{F_{TT}^{\cos(2\phi_V - \phi_{S_A} - \phi_{S_B})}}{F_{UU}},
\\
& A_{TT}^{1} = \frac{F_{TT}^{1}}{F_{UU}},
\qquad
A_{TT}^{2} = \frac{F_{TT}^{2}}{F_{UU}}.
\eea
Likewise, we define $A_{UT}^{\sin(\phi_V - \phi_{S_B})}$, $A_{UT}^{\cos(\phi_V - \phi_{S_B})}$, $A_{LT}^{\sin(\phi_V - \phi_{S_B})}$, and $A_{LT}^{\cos(\phi_V - \phi_{S_B})}$ as the ratios of the corresponding structure functions to the unpolarized structure function $F_{UU}$, respectively. Following the relations in Eqs.~(\ref{eq:asyLU} - \ref{eq:asyTTsin}), for the polarized proton-proton collisions, we have
\bea
&A_{LU}(y) = A_{UL} (-y), 
\qquad
A_{LL}(y) = A_{LL} (-y), 
\label{eq:ALU}
\\
&A_{TU}^{\sin(\phi_V - \phi_{S_A})}(y) = - A_{UT}^{\sin(\phi_V - \phi_{S_B})} (-y),
\qquad
A_{TU}^{\cos(\phi_V - \phi_{S_A})}(y) =  A_{UT}^{\cos(\phi_V - \phi_{S_B})} (-y),
\label{eq:ATU}
\\
&A_{TL}^{\sin(\phi_V - \phi_{S_A})}(y) = - A_{LT}^{\sin(\phi_V - \phi_{S_B})} (-y),
\qquad
A_{TL}^{\cos(\phi_V - \phi_{S_A})}(y) = A_{LT}^{\cos(\phi_V - \phi_{S_B})} (-y).
\label{eq:ATL}
\\
&A_{TT}^{\cos(2\phi_V - \phi_{S_A} - \phi_{S_B})}(y) = A_{TT}^{\cos(2\phi_V - \phi_{S_A} - \phi_{S_B})}(-y),
\qquad
A_{TT}^{1}(y) = A_{TT}^{1}(-y),
\label{eq:ATTcos}
\\
&A_{TT}^{\sin(2\phi_V - \phi_{S_A} - \phi_{S_B})}(y) = - A_{TT}^{\sin(2\phi_V - \phi_{S_A} - \phi_{S_B})}(-y),
\qquad
A_{TT}^{2}(y) = - A_{TT}^{2}(-y).
\label{eq:ATTsin}
\eea
Looking at the expressions for all the structure functions given in last section, we find that $A_{LL}$, $A_{TU}^{\sin(\phi_V - \phi_{S_A})}$, $A_{UT}^{\sin(\phi_V - \phi_{S_B})}$, $A_{TL}^{\cos(\phi_V - \phi_{S_A})}$, $A_{LT}^{\cos(\phi_V - \phi_{S_B})}$, $A_{TT}^{\cos(2\phi_V - \phi_{S_A} - \phi_{S_B})}$, and $A_{TT}^{1}$ exist for both $W^{\pm}/Z^0$ and $\gamma^*$, and they are parity-even spin asymmetries. On the other hand, $A_{LU},~A_{UL}$, $A_{TU}^{\cos(\phi_V - \phi_{S_A})}$, $A_{UT}^{\cos(\phi_V - \phi_{S_B})}$, $A_{TL}^{\sin(\phi_V - \phi_{S_A})}$, $A_{LT}^{\sin(\phi_V - \phi_{S_B})}$, $A_{TT}^{\sin(2\phi_V - \phi_{S_A} - \phi_{S_B})}$, and $A_{TT}^{2}$ are directly proportional to $a_q$ (the axial coupling), and thus they exist because of the parity-violating nature of the weak interaction, i.e. they are all parity-odd spin asymmetries. 

Single transverse spin asymmetries have also been denoted as $A_N$~\cite{Kang:2009bp,Kang:2009sm,Kang:2011hk,Kang:2012xf,Gamberg:2013kla} in the literature, where usually one chooses a frame such that the transversely polarized proton $A$ moves in the $+z$-direction, and the spin vector $\vec S_{AT}$ and the transverse momentum $\vec q_T$ are along the $y$ and $x$ directions, respectively. It is important to realize that such definition of $A_N$ is related to our definition of $A_{TU}^{\sin(\phi_V - \phi_{S_A})}$ by a minus sign~\cite{Kang:2009sm}:
\bea
A_{TU}^{\sin(\phi_V - \phi_{S_A})} = - A_N.
\label{eq-definition}
\eea
It is worth mentioning that $A_N$ (and thus $A_{TU}^{\sin(\phi_V - \phi_{S_A})}$) is usually referred to as ``left-right'' spin asymmetry~\cite{Adams:1991cs,Adams:2003fx,Abelev:2008af,Gamberg:2014eia,Kang:2011hk,Perdekamp:2015vwa}. In this sense, one could refer to the other single transverse spin asymmetry $A_{TU}^{\cos(\phi_V - \phi_{S_A})}$ as ``up-down'' spin asymmetry. 

In this section we will  present some numerical estimates for these spin asymmetries defined above. These spin asymmetries are all expressed in terms of various TMDs, as shown in the last section. In principle, to make precise quantitative predictions, one has to take into account the effect of the TMD evolution. The TMD evolution will likely lead to suppression of these spin asymmetries; see e.g.  Refs.~\cite{Echevarria:2014xaa,Boer:2013zca,Boer:2001he,Aybat:2011ge,Aybat:2011ta,Aidala:2014hva,Sun:2013hua,Anselmino:2012aa}. However, at this point the phenomenological implementation of TMD evolution formalism still has very large uncertainties. In fact, one of the motivations for the transverse spin asymmetry measurements of the vector boson production in polarized proton-proton collisions at RHIC is to constrain the TMD evolution formalism. Because of this, we will only present the numerical estimate of the spin asymmetries using the usual Gaussian model for all the quark TMDs, i.e., without TMD evolution. Since the TMD evolution is supposed to apply to all the TMDs roughly equally, the hope is that the relative magnitude of the various spin asymmetries could serve as a reasonable  guidance for the experimental measurements. 

Both the unpolarized quark TMD $f_1^q$ and the helicity TMD distribution $g_{1L}^q$ are $k_T$-even functions and we assume they have the following Gaussian forms:
\bea
f_1^q(x,k_T^2) &= f_1^q(x)\frac{1}{\pi\langle k_T^2\rangle_{f_1}}e^{-k_T^2/\langle k_T^2\rangle_{f_1}},
\nnu
g_{1L}^q(x,k_T^2) &= g_{1L}^q(x)\frac{1}{\pi \langle k_T^2\rangle_{g_{1L}}}
e^{-k_T^2/\langle k_T^2\rangle_{g_{1L}}},
\eea
where $f_1^q(x)$ and $g_{1L}^q(x)$ are the collinear unpolarized parton distribution function and helicity distribution function, respectively. We will assume the Gaussian width for $f_1^q$ and $g_{1L}^q$ to be the same~\cite{Kang:2015msa,Kang:2014zza}, and take the value of 0.25 GeV$^2$~\cite{Anselmino:2005nn}:
\bea
\langle k_T^2\rangle_{f_1} = \langle k_T^2\rangle_{g_{1L}} = 0.25~ {\rm GeV}^2.
\eea
On the other hand, both the Sivers function $f_{1T}^{\perp q}$ and the transversal helicity distribution $g_{1T}^q$ are the coefficients from the linear $k_T$-expansion term of the quark-quark correlator $\Phi^q$; see Eqs.~\eqref{eq-gamma+} and \eqref{eq-gamma+5}. Quark Sivers functions have been extracted from semi-inclusive deep inelastic scattering. We take the parametrization from Ref.~\cite{Anselmino:2008sga}:
\bea
\frac{k_T}{M}f_{1T}^{\perp q}(x,k_T^2) = - \mathcal{N}_q(x)h(k_T)f_1^q(x, k_T^2),
\eea
where $\mathcal{N}_q(x)$ and $h(k_T)$ are given by
\bea
\mathcal{N}_q(x) &= N_q x^{\alpha_q}(1-x)^{\beta_q}\frac{(\alpha_q+\beta_q)^{(\alpha_q+\beta_q)}}{\alpha_q^{\alpha_q}\beta_q^{\beta_q}},
\nnu
h(k_T) &= \sqrt{2e}\,\frac{k_T}{M_1} e^{ - k_T^2/M_1^2},
\eea
with the parameters $N_q,~\alpha_q, ~\beta_q$, and $M_1$ given in \cite{Anselmino:2008sga}. At the same time, for $g_{1T}^q$, we have
\bea
\frac{1}{2M^2} g_{1T}^q(x, k_T^2) = g_{1T}^{q(1)}(x)\frac{1}{\pi\langle k_T^2\rangle_{g_{1T}}^2}
e^{-k_T^2/\langle k_T^2\rangle_{g_{1T}}},
\eea
where we choose $\langle k_T^2\rangle_{g_{1T}} = 0.15$ GeV$^2$~\cite{Kotzinian:2006dw}, and $g_{1T}^{q(1)}(x)$ is the first $k_T$ moment of $g_{1T}^q(x, k_T^2)$ defined as
\bea
g_{1T}^{q(1)}(x) = \int d^2k_T\frac{\vec k_T^2}{2M^2} g_{1T}^q(x, k_T^2).
\eea
We note that there is a Wandzura-Wilczek-type approximation for $g_{1T}^{q(1)}(x)$, which relates $g_{1T}^{q(1)}(x)$ to the collinear helicity distribution function $g_{1L}^q(x)$~\cite{Metz:2008ib}:
\bea
g_{1T}^{q(1)}(x) \approx x \int_x^1\frac{dz}{z} g_{1L}^q(z).
\eea
Now the parametrizations for all the TMDs involved in our calculations, $f_1^q, ~f_{1T}^{\perp q},~g_{1L}^q$, and $g_{1T}^q$, are given. One last thing one should keep in mind is that the quark Sivers function changes sign when probed in SIDIS and DY processes~\cite{Collins:2002kn,Boer:2003cm,Brodsky:2002cx,Brodsky:2002rv,Kang:2009bp}:
\bea
f_{1T}^{\perp q}(x, k_T^2)|_{\rm DY/W/Z} = - f_{1T}^{\perp q}(x, k_T^2)|_{\rm SIDIS},
\eea
whereas all other quark TMDs ($f_1^q,~g_{1L}^q,~g_{1T}^q$) are universal in SIDIS and Drell-Yan. As we have mentioned already in the Introduction, since $g_{1T}^{q}(x, k_T^2)$ was investigated in SIDIS measurements~\cite{Parsamyan:2007ju,Pappalardo:2011cu,Huang:2011bc,Parsamyan:2015dfa} and is proposed to be measured with high precision in future SIDIS experiments~\cite{Gao:2010av}, one could in principle  test the universality of $g_{1T}^{q}(x, k_T^2)$ in $W^{\pm}/Z^0$ production in proton-proton collisions at the RHIC. In our numerical studies below, we implement the sign change for the quark Sivers function and keep all other TMDs the same as in SIDIS when we present the asymmetries of $W^{\pm}/Z^{0}$ boson production at RHIC, to which we now turn.

%%%%%%%%%%%%%%%%%
\subsection{Numerical estimate for the spin asymmetries}
We now present the numerical estimates for the spin asymmetries. We use CTEQ6~\cite{Pumplin:2002vw} for the collinear unpolarized parton distribution functions $f_{1}^{q}(x)$, DSSV~\cite{deFlorian:2008mr,deFlorian:2009vb} parametrization for the collinear helicity distribution functions $g_{1L}^{q}(x)$, and choose the factorization scale $\mu=M_V$. Within our approximation, where $\langle k_T^2\rangle_{f_1} = \langle k_T^2\rangle_{g_{1L}}$, there is no $q_T$ dependence for the longitudinal spin asymmetries $A_{LU}, ~A_{UL}, ~A_{LL}$. This is because $F_{LU}, ~F_{UL}, ~F_{LL}$ all have the same $q_T$ dependence as the unpolarized structure function $F_{UU}$, which then cancels out in the longitudinal spin asymmetries. In this case, these longitudinal spin asymmetries will be the same as those for the inclusive ($q_T$-integrated) vector boson production, which are available in the literature, see e.g. Refs.~\cite{Bourrely:1993dd,Bourrely:1994sc,Gehrmann:1997ez,deFlorian:2010aa,vonArx:2011fz,Kang:2011qz}. We will, thus, not present numerical estimates for these longitudinal spin asymmetries here. Instead, we will focus on those spin asymmetries which involve transversely polarized proton in the collisions, where one has to measure the azimuthal angle of the vector boson in order to probe these asymmetries. In particular, we will present the numerical results for $A_{TU}^{\sin(\phi_V-\phi_{S_A})}$, $A_{TU}^{\cos(\phi_V-\phi_{S_A})}$, $A_{TL}^{\sin(\phi_V-\phi_{S_A})}$, $A_{TL}^{\cos(\phi_V-\phi_{S_A})}$, $A_{TT}^{\sin(2\phi_V-\phi_{S_A}-\phi_{S_B})}$ and $A_{TT}^{\cos(2\phi_V-\phi_{S_A}-\phi_{S_B})}$. The spin asymmetries involving the transverse spin vector $\vec S_{BT}$ (i.e. $\phi_{S_B}$) can be obtained from the results we present through the relations established in Eqs.~\eqref{eq:ATU} and \eqref{eq:ATL}. 
\bef
\includegraphics[width=3.04in]{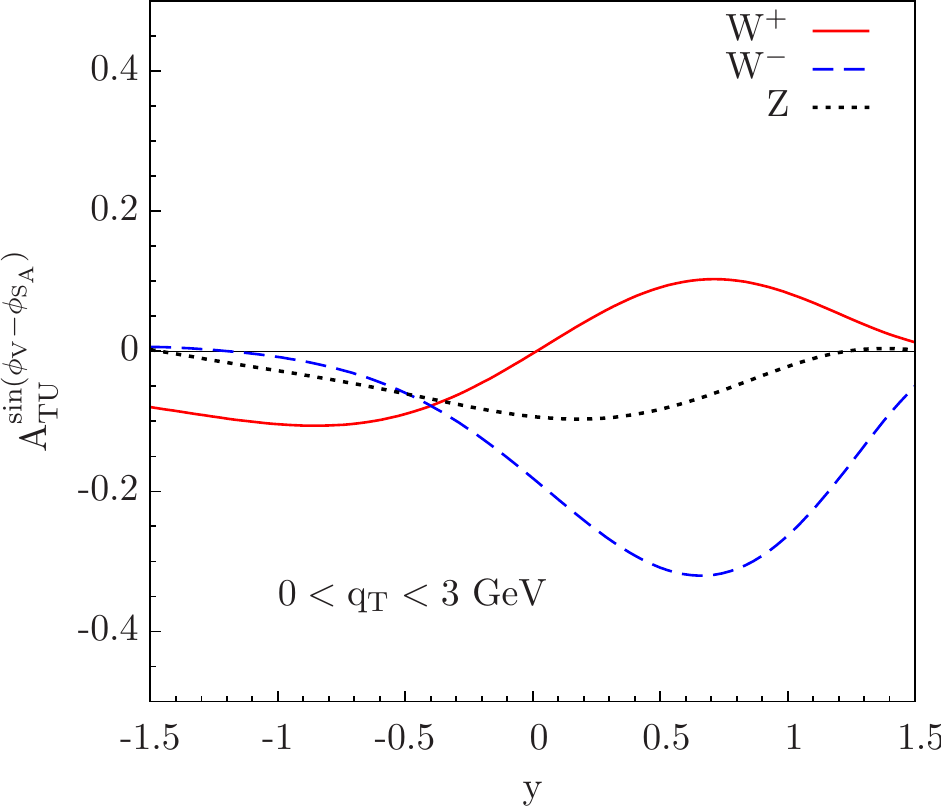}
\hskip 0.2in
\includegraphics[width=3.0in]{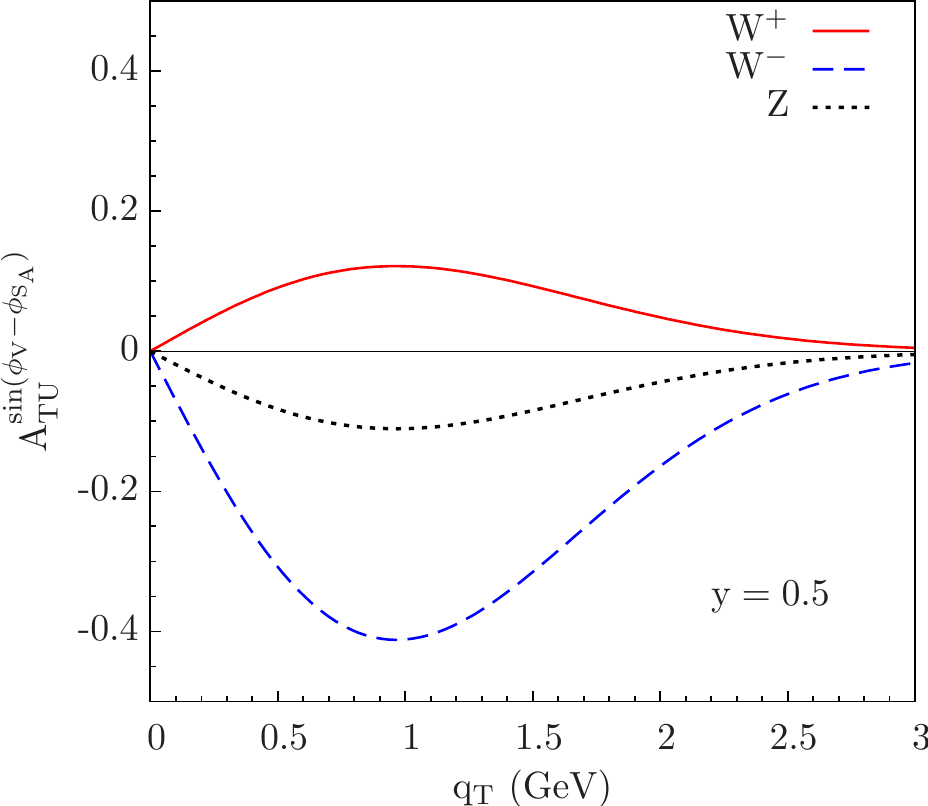}
\caption{Single transverse spin asymmetry $A_{TU}^{\sin(\phi_V-\phi_{S_A})}$ as a function of the rapidity $y$ of the vector boson (left), and as a function of the transverse momentum $q_T$ of the vector boson at rapidity $y=0.5$ (right) at the RHIC energy $\sqrt{s} = 510$ GeV. In the left plot, we have integrated vector boson transverse momentum in the region $0 < q_T < 3$ GeV. The red solid curve is for $W^+$, the blue dashed curve is for $W^-$, and the black dotted curve is for $Z^0$ production. Note: $A_{TU}^{\sin(\phi_V-\phi_{S_A})}$ is related to the parity-conserving interaction (parity-even), and can be used to probe the quark Sivers function $f_{1T}^{\perp q}(x, k_T^2)$.}
\label{ATUsin}
\eef
\bef
\includegraphics[width=3.04in]{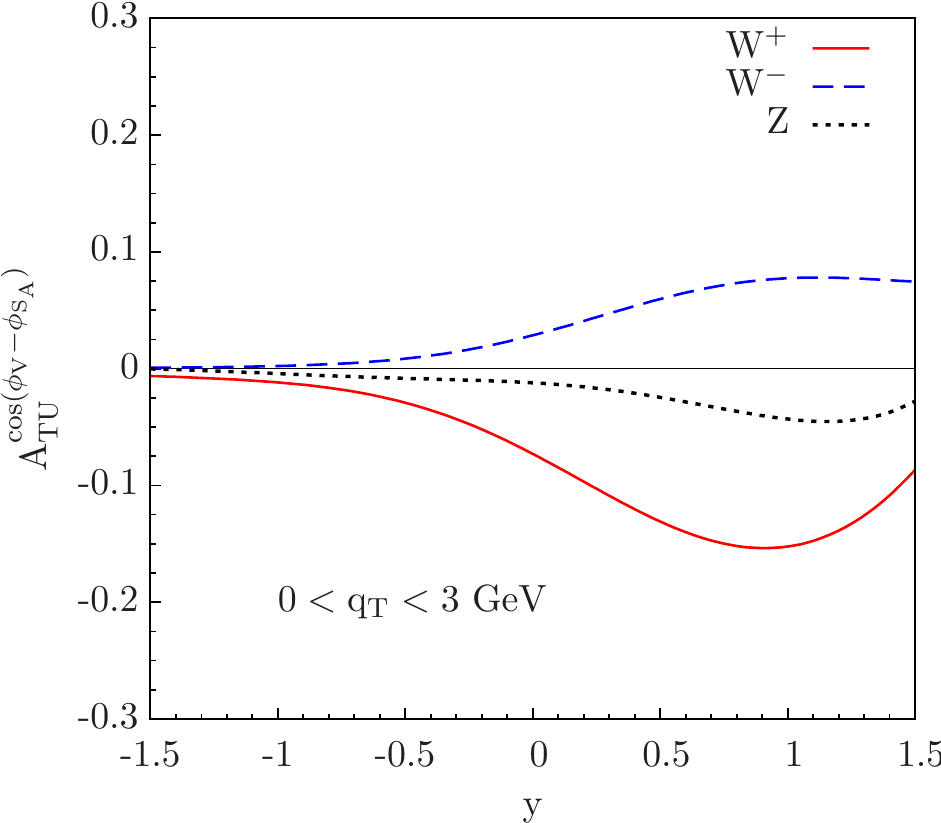}
\hskip 0.2 in
\includegraphics[width=3.0in]{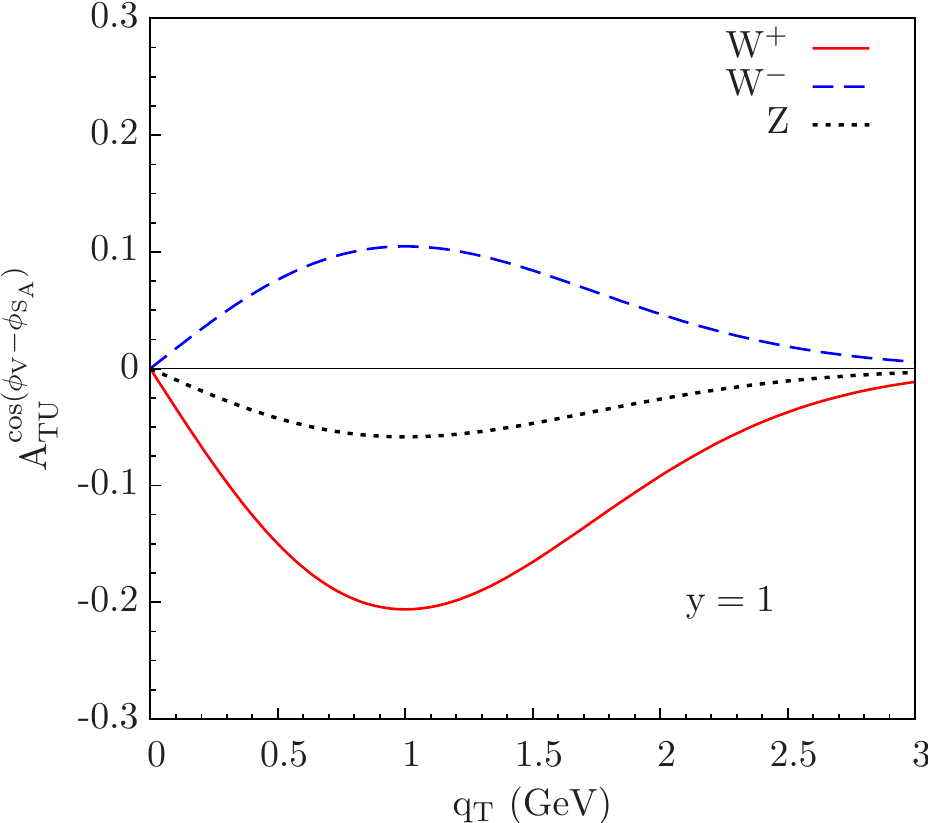}
\caption{Single transverse spin asymmetry $A_{TU}^{\cos(\phi_V-\phi_{S_A})}$ as a function of the rapidity $y$ of the vector boson (left), and as a function of the transverse momentum $q_T$ of the vector boson at rapidity $y=1$ (right) at the RHIC energy $\sqrt{s} = 510$ GeV. In the left plot, we have integrated vector boson transverse momentum in the range $0 < q_T < 3$ GeV. The red solid curve is for $W^+$, the blue dashed curve is for $W^-$, and the black dotted curve is for $Z^0$ production. Note: $A_{TU}^{\cos(\phi_V-\phi_{S_A})}$ is related to the parity-violating interaction (parity-odd), and can be used to probe the  transversal helicity distribution $g_{1T}^{q}(x, k_T^2)$.}
\label{ATUcos}
\eef

Let us first discuss the single transverse spin asymmetry: $A_{TU}^{\sin(\phi_V-\phi_{S_A})}$ and $A_{TU}^{\cos(\phi_V-\phi_{S_A})}$. In Fig.~\ref{ATUsin}, we plot single transverse spin asymmetry $A_{TU}^{\sin(\phi_V-\phi_{S_A})}$ as a function of the rapidity $y$ of the vector boson (left), and as a function of the transverse momentum $q_T$ of the vector boson at rapidity $y=0.5$ (right) at the RHIC energy $\sqrt{s} = 510$ GeV. In the left plot, we have integrated the vector boson transverse momentum in the region $0 < q_T < 3$ GeV. The red solid curve is for $W^+$, the blue dashed curve is for $W^-$, and the black dotted curve is for $Z^0$ production. These results are consistent with those in \cite{Kang:2009bp} and \cite{Kang:2009sm}, respectively~\footnote{Of course one has to keep in mind that $A_N$ were plotted in \cite{Kang:2009bp} and \cite{Kang:2009sm}, while here $A_{TU}^{\sin(\phi_V-\phi_{S_A})}$ are plotted. So they should have an opposite sign because of Eq.~\eqref{eq-definition}.}. It is worthwhile to remind the reader that $A_{TU}^{\sin(\phi_V-\phi_{S_A})}$ is parity-even, and can be used to probe the quark Sivers function $f_{1T}^{\perp q}(x, k_T^2)$. Measuring such a transverse spin asymmetry and testing the sign change of the quark Sivers function is one of the main goals of the transverse $W$ program in the near future at RHIC~\cite{Aschenauer:2015eha}. One should also keep in mind that because the sea quark Sivers functions are not really constrained from the fixed-target SIDIS measurements, our theoretical curves could have very large uncertainties~\cite{Adamczyk:2015gyk}, especially in the backward rapidity region where the asymmetry is most sensitive to the sea quark distributions. 
\bef
\includegraphics[width=3.04in]{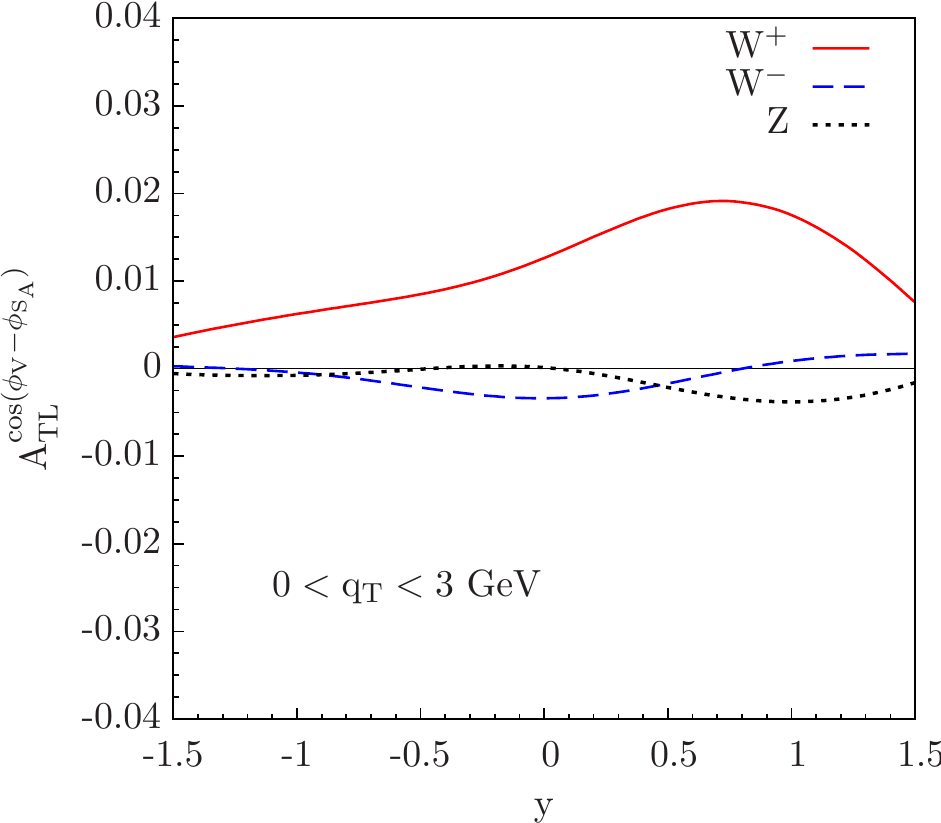}
\hskip 0.2in
\includegraphics[width=3.0in]{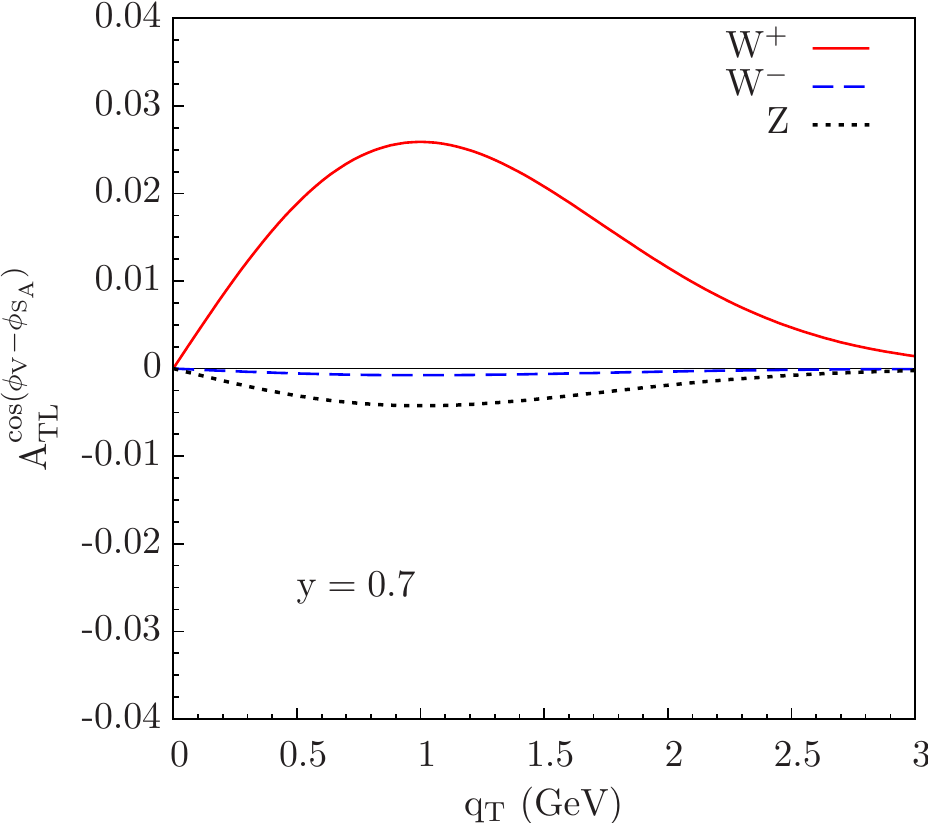}
\caption{Transverse-longitudinal double spin asymmetries $A_{TL}^{\cos(\phi_V-\phi_{S_A})}$ as a function of rapidity $y$ (left) and as a function of transverse momentum $q_T$ of the vector boson at rapidity $y=0.7$ (right) at the RHIC energy $\sqrt{s} = 510$ GeV.}
\label{ATLcos}
\eef
\bef
\includegraphics[width=3.04in]{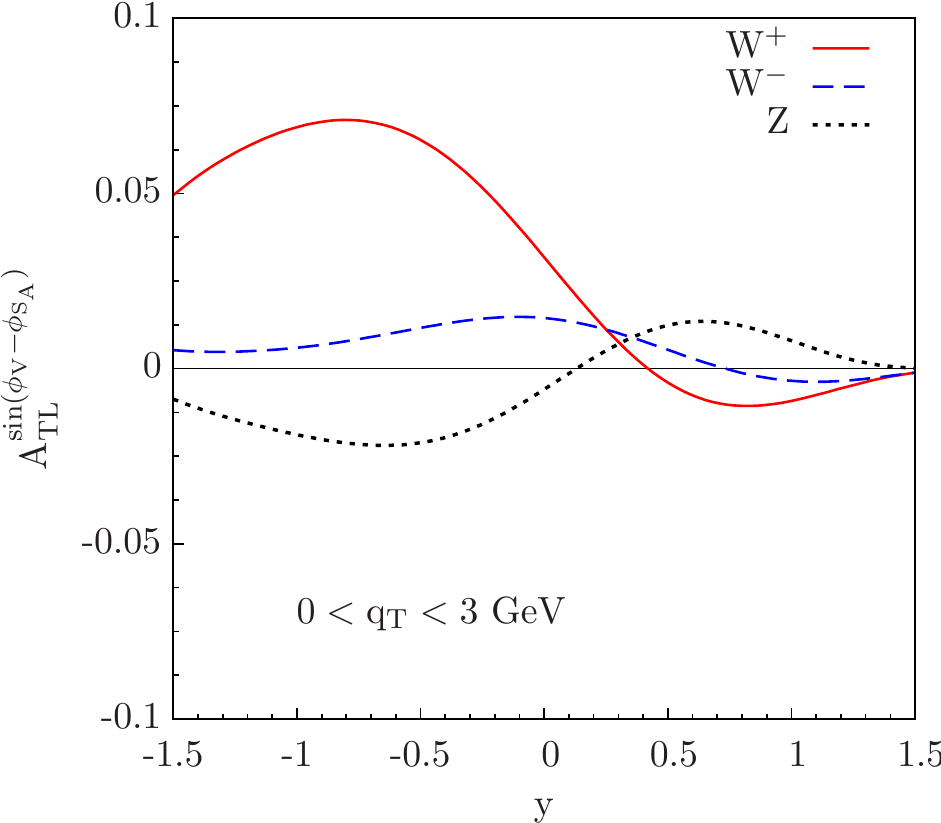}
\hskip 0.2in
\includegraphics[width=3.0in]{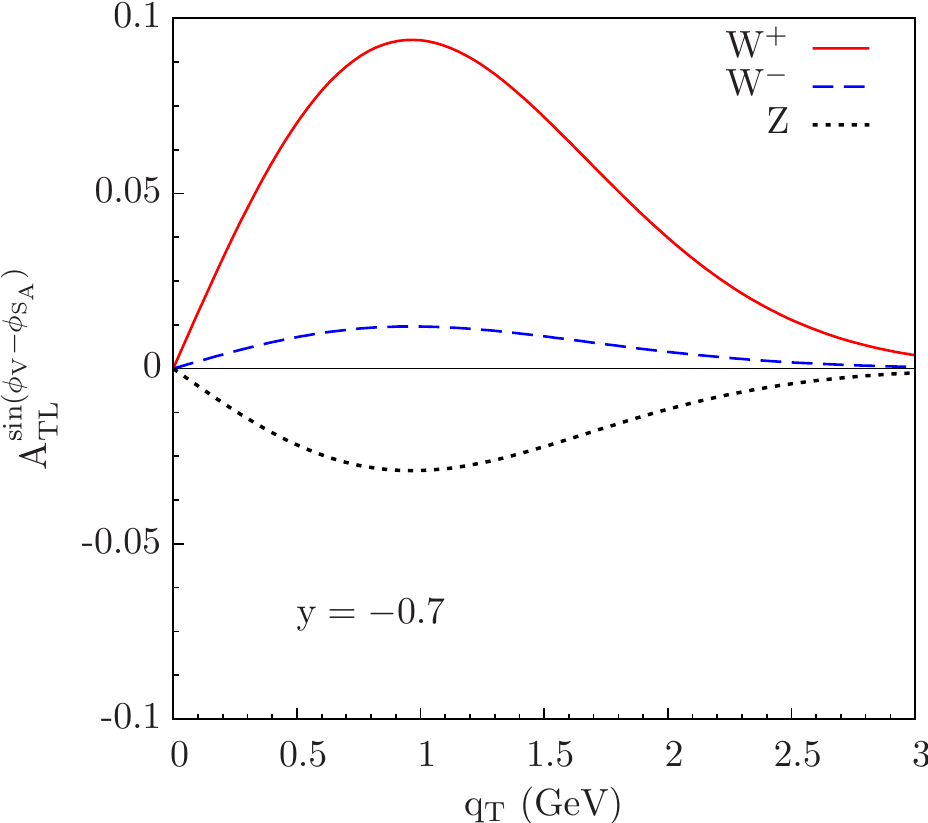}
\caption{Transverse-longitudinal spin asymmetries double spin asymmetries $A_{TL}^{\sin(\phi_V-\phi_{S_A})}$ as a function of rapidity $y$ (left) and as a function of transverse momentum $q_T$ of the vector boson at rapidity $y=-0.7$ (right) at the RHIC energy $\sqrt{s} = 510$ GeV.}
\label{ATLsin}
\eef

On the other hand, due to the parity-violating interaction, there is another single transverse spin asymmetry, $A_{TU}^{\cos(\phi_V-\phi_{S_A})}$ (parity odd). This term is related to $g_{1T}^{q}(x, k_T^2)$. Since $g_{1T}^{q}$ is not fully constrained even for the valence quarks within the current SIDIS measurements~\cite{Parsamyan:2007ju,Pappalardo:2011cu,Huang:2011bc,Parsamyan:2015dfa}, $A_{TU}^{\cos(\phi_V-\phi_{S_A})}$ of $W^{\pm}/Z^0$ production at RHIC could on one hand serve as a complementary channel to constrain $g_{1T}^{q}$, at the same time, as we have emphasized, one in principle could test the universality of $g_{1T}^{q}$ as to the future high precision SIDIS measurement~\cite{Gao:2010av,Accardi:2012qut}. In Fig.~\ref{ATUcos}, we plot $A_{TU}^{\cos(\phi_V-\phi_{S_A})}$ as a function of the rapidity $y$ of the vector boson (left), and as a function of the transverse momentum $q_T$ of the vector boson at rapidity $y=1$ (right) at the RHIC energy $\sqrt{s} = 510$ GeV. 
The asymmetry is sizable, in particular due to the fact that $W^{\pm}$ production provides maximum analyzing power for the quark longitudinal polarization. If TMD evolution only leads to a moderate suppression, this asymmetry should be measurable at the RHIC. 
\bef
\includegraphics[width=3.04in]{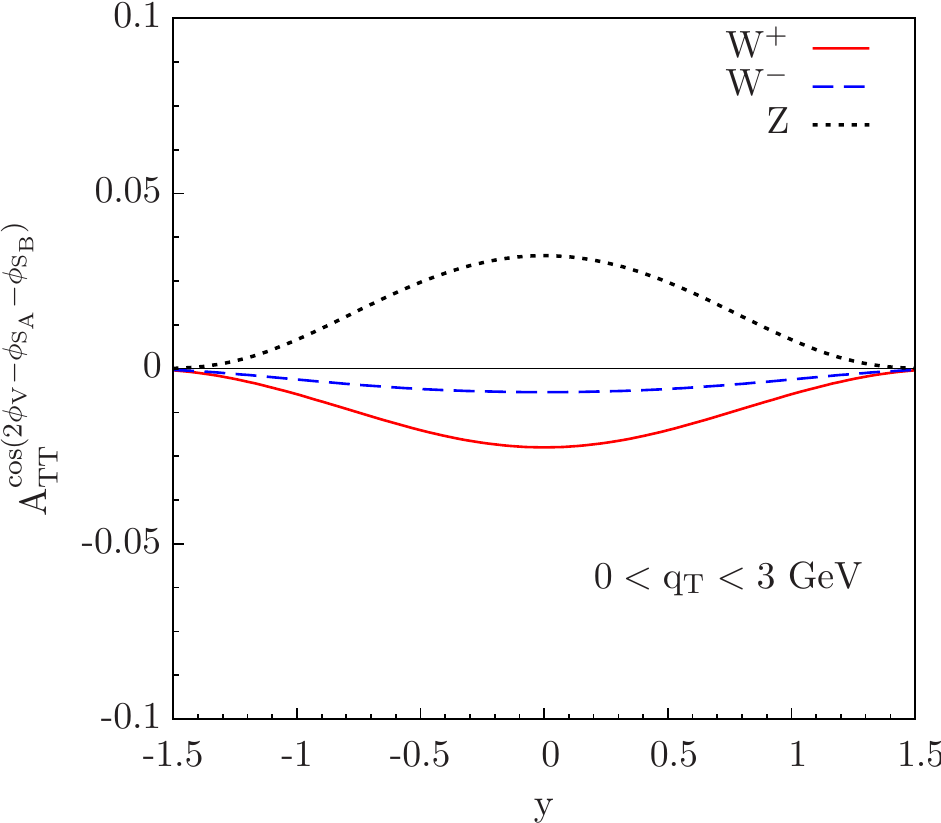}
\hskip 0.2in
\includegraphics[width=3.0in]{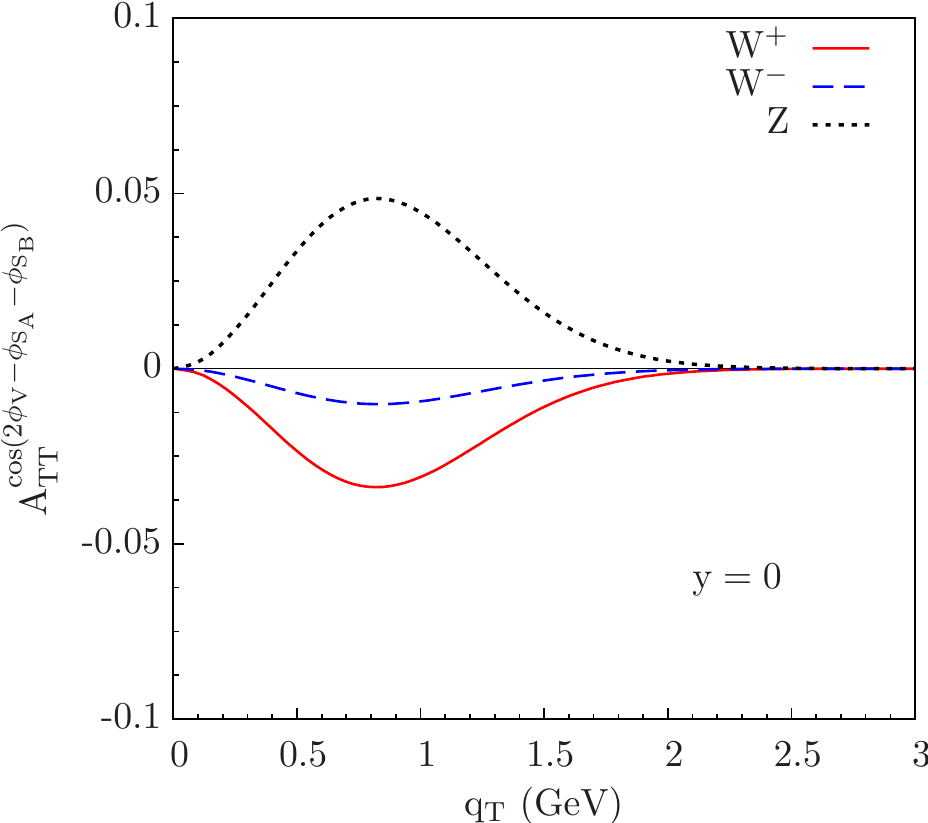}
\caption{Double transverse spin asymmetry $A_{TT}^{\cos(2\phi_V-\phi_{S_A}-\phi_{S_B})}$ as a function of rapidity $y$ (left) and as a function of transverse momentum $q_T$ of the vector boson at rapidity $y=0$ (right) at the RHIC energy $\sqrt{s} = 510$ GeV.}
\label{ATTcos}
\eef

Let us now turn to the study of the transverse-longitudinal double spin asymmetry $A_{TL}^{\cos(\phi_V-\phi_{S_A})}$ and $A_{TL}^{\sin(\phi_V-\phi_{S_A})}$. While $A_{TL}^{\cos(\phi_V-\phi_{S_A})}$ is related to the parity-conserving interaction and is sensitive to both $g_{1T}^q$ and $g_{1L}^q$, $A_{TL}^{\sin(\phi_V-\phi_{S_A})}$ is related to the parity-violating interaction and is sensitive to $f_{1T}^{\perp q}$ and $g_{1L}^q$. In Fig.~\ref{ATLcos}, we plot the transverse-longitudinal double spin asymmetries $A_{TL}^{\cos(\phi_V-\phi_{S_A})}$ as a function of rapidity $y$ (left) and as a function of the transverse momentum $q_T$ of the vector boson at rapidity $y=0.7$ (right). Since both $g_{1T}^q$ and $g_{1L}^q$ are suppressed when compared with the unpolarized parton distribution $f_1^q$, the double spin asymmetry is indeed much smaller than the single transverse spin asymmetry, where only one spin-dependent parton distribution (either $f_{1T}^{\perp q}$ or $g_{1T}^q$) is involved. In Fig.~\ref{ATLsin}, we plot $A_{TL}^{\sin(\phi_V-\phi_{S_A})}$ as a function of rapidity $y$ (left) and as a function of transverse momentum $q_T$ of the vector boson at rapidity $y=-0.7$ (right) at the top RHIC energy. Even though the asymmetry is in general not very large, $A_{TL}^{\sin(\phi_V-\phi_{S_A})}$ for $W^+$ is quite sizable, $\sim 5\%-10\%$. The reason lies in the fact that the Sivers function $f_{1T}^{\perp q}$ from Ref.~\cite{Anselmino:2008sga} for $\bar d$ quark is still sizable and $g_{1L}^{q}$ for $u$ quark is reasonably large. Thus, their product $f_{1T}^{\perp \bar d} \, g_{1L}^{u}$ leads to a large double spin asymmetry.
\bef
\includegraphics[width=3.04in]{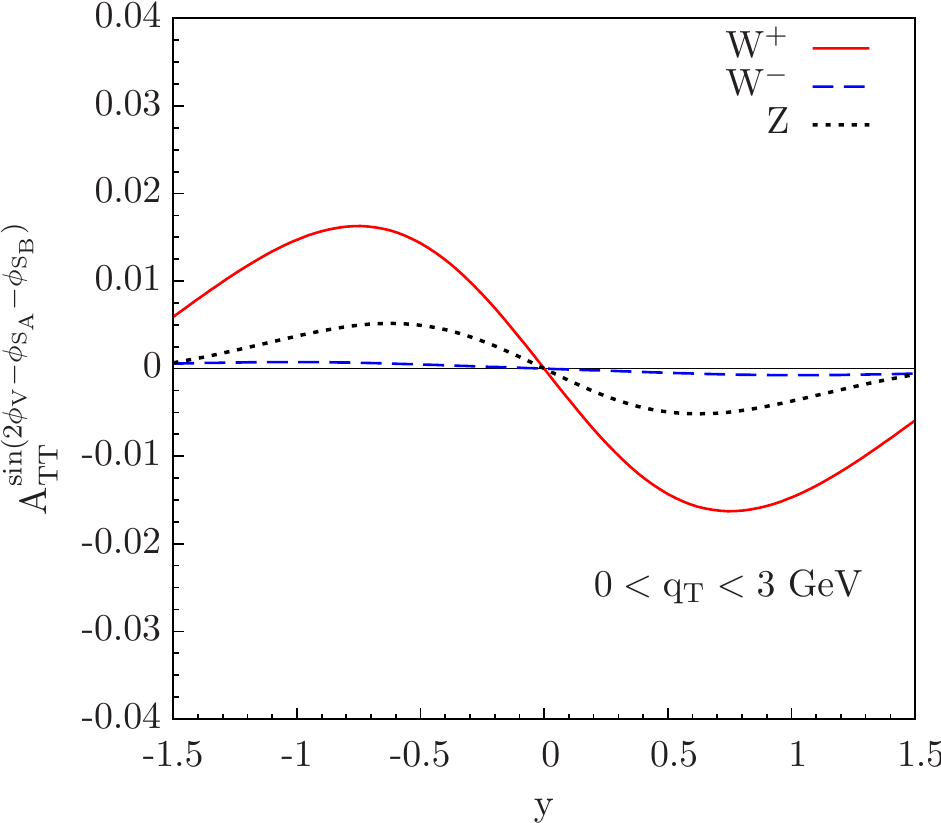}
\hskip 0.2in
\includegraphics[width=3.0in]{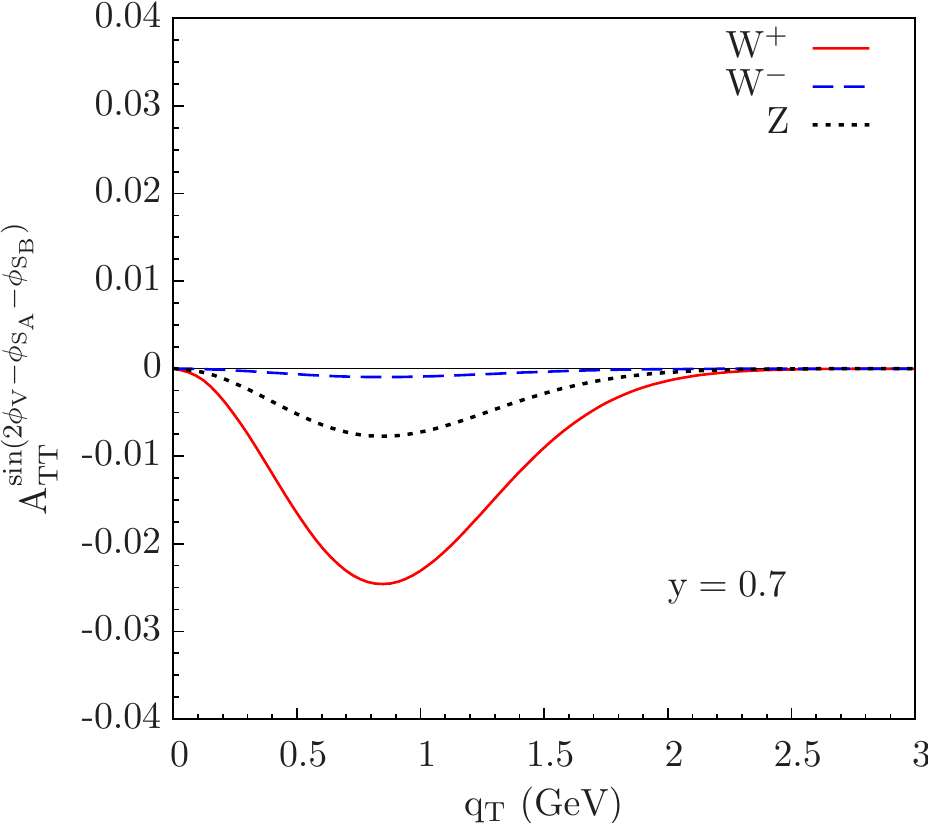}
\caption{Double transverse spin asymmetry $A_{TT}^{\sin(2\phi_V-\phi_{S_A}-\phi_{S_B})}$ as a function of rapidity $y$ (left) and as a function of transverse momentum $q_T$ of the vector boson at rapidity $y=0.7$ (right) at the RHIC energy $\sqrt{s} = 510$ GeV.}
\label{ATTsin}
\eef

Finally, we study the double transverse spin asymmetries. As examples, we present the numerical results for $A_{TT}^{\cos(2\phi_V-\phi_{S_A}-\phi_{S_B})}$ and $A_{TT}^{\sin(2\phi_V-\phi_{S_A}-\phi_{S_B})}$, which still involve the azimuthal angle $\phi_V$ of the vector boson (the other two asymmetries $A_{TT}^{1}$ and $A_{TT}^{2}$ do not). In Fig.~\ref{ATTcos} we plot $A_{TT}^{\cos(2\phi_V-\phi_{S_A}-\phi_{S_B})}$ as a function of rapidity $y$ (left) and as a function of transverse momentum $q_T$ of the vector boson at rapidity $y=0.7$ (right) at the RHIC energy $\sqrt{s} = 510$ GeV. Only the $W^+$ and the $Z^0$ bosons have a reasonable large asymmetries $\lesssim 5\%$ but with opposite sign. This can be understood as follows. We have $f_{1T}^{\perp u} f_{1T}^{\perp \bar d}$ contributing to $W^+$ while $f_{1T}^{\perp d} f_{1T}^{\perp \bar d}$ to the $Z$ boson. From the Sivers function parametrization we used \cite{Anselmino:2008sga}, the $u$ and $d$ quark Sivers distributions have the opposite sign. Together with the fact that $f_{1T}^{\perp \bar d}$ is still sizable in this parametrization, this leads to opposite but reasonably large asymmetry for $W^+$ and $Z^0$ bosons. In Fig.~\ref{ATTsin} we plot $A_{TT}^{\sin(2\phi_V-\phi_{S_A}-\phi_{S_B})}$ as a function of rapidity $y$ (left) and as a function of transverse momentum $q_T$ of the vector boson at rapidity $y=0.7$ (right) at the RHIC energy $\sqrt{s} = 510$ GeV. The slightly larger asymmetries $A_{TT}^{\sin(2\phi_V-\phi_{S_A}-\phi_{S_B})}$ for $W^+/Z^0$ can be understood similarly. 

%%%%%%%%%%%%%%
\section{Summary}
\label{sec:IV}

In this paper we studied the spin-dependent differential cross sections for  vector boson ($W^{\pm}/Z^0/\gamma^*$) production in polarized nucleon-nucleon collisions for low transverse momentum of the observed vector boson. We considered the situation where the full kinematics of the vector boson could be reconstructed and both the magnitude $q_T$ and azimuthal angle $\phi_V$ of the vector bosons are measured. We presented the cross sections and the associated single and double spin asymmetries in terms of the transverse momentum dependent parton distribution functions (TMDs) at tree level within the TMD factorization formalism. We estimated these spin asymmetries for $W^{\pm}/Z^0$ boson production in polarized proton-proton collisions at the top RHIC energy, and found that if the TMD evolution effect does not lead to too strong a suppression, some of the asymmetries are rather sizable and should be measurable at RHIC. 
In particular, the single transverse spin asymmetries contain two large orthogonal azimuthal terms: a parity-conserving term that is  sensitive to the quark Sivers function $f_{1T}^{\perp q}$, as well as a parity-violating term that probes the quark transversal helicity distribution $g_{1T}^{q}$. While $f_{1T}^{\perp q}$ is predicted to change sign from semi-inclusive deep inelastic scattering to the Drell-Yan process, $g_{1T}^{q}$ is supposed to be universal. Thus, the $W$ spin physics  program at RHIC could be viewed as truly multipurpose: one that tests the universality properties of TMDs, constrains the TMD evolution effects, and  probes the sea quark TMDs. 

\section*{Acknowledgments}
We thank Elke-Caroline Aschenauer for helpful comments. This work is supported by the U.S. Department of Energy under Contract Nos.~DE-SC0012704 (J.H.) and DE-AC52-06NA25396 (Z.K., I.V. and H.X.), and in part by the LDRD program at LANL.

\end{document}